\documentclass[aps,prb,superscriptaddress,floats,twocolumn,showpacs,eqsecnum]{revtex4}

\usepackage{verbatim}
\usepackage{amsmath}
\usepackage{amsfonts}
\usepackage{amssymb}
\usepackage{graphicx}
\usepackage{bm}
\usepackage{color}
\usepackage{mathrsfs}
\usepackage{epsf}
\usepackage[hypertex]{hyperref}

\begin{document}
\title{
Boundary criticality at the Anderson transition between a metal and \\ 
a quantum spin Hall insulator in two dimensions
      }

\author{Hideaki Obuse}
\altaffiliation{Present address: Department of Physics, Kyoto University,
Kyoto 606-8502, Japan.}
\affiliation{Condensed Matter Theory Laboratory, RIKEN, Wako,
Saitama 351-0198, Japan}
\author{Akira Furusaki}
\affiliation{Condensed Matter Theory Laboratory, RIKEN, Wako,
Saitama 351-0198, Japan}
\author{Shinsei Ryu}
\affiliation{Kavli Institute for Theoretical Physics, University of
California, Santa Barbara, California 93106, USA}
\author{Christopher Mudry}
\affiliation{Condensed Matter Theory Group, Paul Scherrer Institute,
CH-5232 Villigen PSI, Switzerland}

\begin{abstract}
Static disorder in a noninteracting gas of electrons confined
to two dimensions can drive a continuous quantum (Anderson) transition
between a metallic and an insulating state when time-reversal symmetry 
is preserved but spin-rotation symmetry is broken.
The critical exponent $\nu$ that characterizes the
diverging localization length and the bulk 
multifractal scaling exponents that characterize the
amplitudes of the critical wave functions at the
metal-insulator transition do not depend on the topological nature
of the insulating state, i.e., whether it is topologically 
trivial (ordinary insulator) or nontrivial (a $\mathbb{Z}^{\ }_2$ 
insulator supporting a quantum spin Hall effect). This is not true
of the boundary multifractal scaling exponents which we show (numerically)
to depend on whether the insulating state is topologically
trivial or not.
\end{abstract}

\pacs{73.20.Fz, 71.70.Ej, 73.43.-f, 05.45.Df}

\date{\today}

\maketitle

\section{
Introduction
        }

It has long been known that the metallic state
of a two-dimensional gas of noninteracting electrons 
is robust to sufficiently weak static disorder 
when time-reversal symmetry (TRS) is preserved 
but spin-rotation symmetry (SRS) is broken.%
~\cite{Hikami80}
Increasing the disorder strength relative to the Fermi energy
induces a continuous quantum (Anderson) transition to an insulating
state. Bulk properties of this metal-insulator transition
have been well characterized numerically
when the insulating state is topologically trivial.%
~\cite{Asada02,Obuse07a,Mildenberger07a}

It has only been realized in the last few years that
there are two distinct classes of 
time-reversal invariant (band) insulators:
$\mathbb{Z}^{\ }_{2}$ topologically trivial and nontrivial insulators.
The $\mathbb{Z}^{\ }_{2}$ topological insulator in two dimensions
supports a helical edge state
which is a Kramers pair of counter propagating gapless excitations.%
~\cite{Kane05,Bernevig06}
This helical edge state is responsible for 
the quantum spin Hall (QSH) effect:
an electric field induces a spin accumulation
on the edges transverse to the direction of the electric field.%
~\cite{Kane05,Bernevig06}
The QSH effect has been observed
in HgTe/(Hg,Cd)Te quantum wells.\cite{Bernevig06b,Konig07}
Just as edge states in the integer quantum Hall effect 
are stable against disorder,
the helical edge state in a $\mathbb{Z}^{\ }_{2}$ topological insulator
survives impurity scattering as long as
the bulk energy gap is open and the TRS is preserved.\cite{Wu,Xu}
This implies that, even in the presence of disorder, time-reversal
invariant insulators can be classified into two distinct classes,
$\mathbb{Z}^{\ }_{2}$ topological and nontopological (ordinary) insulators,
according to the presence or absence of a helical edge state.

It is then natural to ask~\cite{Onoda07}
whether the transition between the metallic and the QSH insulating state
belongs to a universality class different than that of 
the (ordinary) two-dimensional symplectic universality
class discovered in Ref.~\onlinecite{Hikami80}.
It was shown in Ref.%
~\onlinecite{Obuse07b}
that the answer is negative for the scaling exponent $\nu$
of the diverging localization length upon approaching the
Anderson transition from the insulating sides.
In this paper, we show numerically that there are 
boundary multifractal scaling exponents \cite{Subramaniam06}
that are sensitive to the
presence or absence of a helical edge state
on the insulating side of the transition.

We review in Sec.~\ref{sec: Network model}
the definition of the two-dimensional 
network model introduced in Ref.~\onlinecite{Obuse07b}
that encodes the transition between the metallic and QSH insulating state,
as well as the transition between the metallic and ordinary insulating state
in the two-dimensional symplectic universality class.
The latter transition is conventionally studied using
the two-dimensional tight-binding model introduced in 
Ref.~\onlinecite{Asada02}, which is also briefly reviewed.
The phase diagram for the network model is reviewed, and
the relevance of boundary conditions to the presence
or absence of helical edge states is discussed
in Sec.~\ref{sec: Phase diagram}.
The dependence of the localization length 
on transverse boundary conditions in quasi-one-dimensional geometries is
discussed in Sec.~\ref{sec: Localization length}.
Boundary multifractal scaling exponents in the network model
are calculated numerically in
Sec.~\ref{sec: Bulk and boundary criticality}.
Corner multifractal scaling exponents 
in the network model are investigated analytically 
in Sec.~\ref{sec:Corner multifractality}.
We conclude with Sec.~\ref{sec: Conclusions}.

\begin{figure}[t]
\centering
\includegraphics[width=0.45\textwidth]{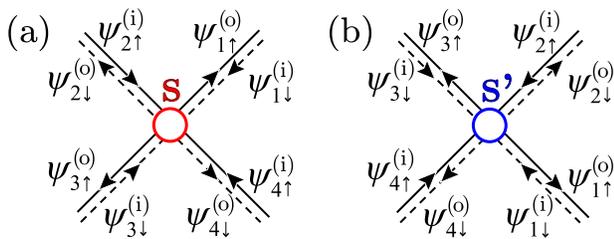}
\caption{
(Color online)
Elementary building blocks of the network model.
A square Bravais lattice with nearest-neighbor sites connected
by bonds underlies the construction of the network model.
Any red site of one of the two sublattices of the square lattice
is replaced by a red circle $\textsf{S}$ 
(a node of type $\textsf{S}$)
with the four bonds meeting at the site
replaced by four pairs of directed links numbered according to the
rule shown in (a). Any blue site of the complementary sublattice
of the square lattice is replaced by a blue circle $\textsf{S}'$
(a node of type $\textsf{S}'$)
with the four bonds meeting at the 
site replaced by four pairs 
of directed links numbered according to the
rule shown in (b). Observe that a clockwise rotation by $\pi/2$
turn (a) into (b).
The directed links represent incoming
or outgoing plane waves with a well-defined
projection of the spin-$1/2$ quantum number along the
quantization axis. Each pair of links replacing a bond
represents a Kramers doublet of plane waves. Each node
$\textsf{S}$ or $\textsf{S}'$ 
depicts a scattering process represented by a $4\times4$
unitary matrix defined in Eq.~(\ref{eq: para S matrix})
that preserves time-reversal symmetry (TRS) but breaks
spin-rotation symmetry (SRS). 
        } 
\label{fig:nodeS}
\end{figure}

\section{
Network model
        }
\label{sec: Network model}

Our starting point is a network model introduced in Ref.%
~\onlinecite{Obuse07b}
to capture the Anderson transition between
the two-dimensional metallic and topological insulating states.
The network model is constructed by decorating an underlying
square lattice of sites and single bonds connecting nearest-neighbor
sites with the elementary building blocks from 
Fig.~\ref{fig:nodeS}. 
By taking advantage of the bipartite nature of 
the square lattice, one colors all sites from one sublattice in red 
and all sites from the complementary sublattice in blue. 
One then replaces any red (blue) site with a node $\textsf{S}$ ($\textsf{S}'$) 
represented graphically by a red (blue) open circle.
Second, any single bond connecting a pair of nearest-neighbor sites 
of the square lattice is replaced by a pair of directed links of 
opposite orientation. On the links, the spin-$1/2$ is a good quantum
number. A link represented by a full line carries
the spin-1/2 quantum number $\sigma=\uparrow$.
A link represented by a dashed line carries
the spin-1/2 quantum number $\sigma=\downarrow$.
Third, the four pairs of directed links that meet at a node
are labeled according to the rules of 
Fig.~\ref{fig:nodeS}(a)
and 
Fig.~\ref{fig:nodeS}(b)
if the node is of type $\textsf{S}$ and $\textsf{S}'$, respectively.
With the conventions of Figs.~\ref{fig:nodeS}(a) and~\ref{fig:nodeS}(b)
either node defines
a $4\times4$
scattering matrix $S$ that preserves TRS but breaks the SRS
and can be represented by
\begin{equation}
\begin{split}
&
\begin{pmatrix}
\psi^{(\mathrm{o})}_{1\uparrow}
\\
\psi^{(\mathrm{o})}_{2\downarrow}
\\
\psi^{(\mathrm{o})}_{3\uparrow}
\\
\psi^{(\mathrm{o})}_{4\downarrow}
\end{pmatrix}
=
S
\begin{pmatrix}
\psi^{(\mathrm{i})}_{2\uparrow}
\\
\psi^{(\mathrm{i})}_{1\downarrow}
\\
\psi^{(\mathrm{i})}_{4\uparrow}
\\
\psi^{(\mathrm{i})}_{3\downarrow}
\end{pmatrix},
\qquad
S=
\left(
\begin{array}{cc}
 r \sigma^{\ }_0 
& 
t Q 
\\
-t Q^{\dag} 
& 
r \sigma^{\ }_0
\end{array}
\right),
\\&
r=\tanh{X},
\qquad
t=\frac{1}{\cosh{X}},
\\
&
Q=
{i}
 \sigma^{\ }_0 \cos\theta \sin\varphi^{\ }_{1}
+
 \sigma^{\ }_1 \sin\theta \cos\varphi^{\ }_{2}
\\
&\hphantom{Q=}
-
 \sigma^{\ }_2 \sin\theta \sin\varphi^{\ }_{2}
+
 \sigma^{\ }_3 \cos\theta \cos\varphi^{\ }_{1}.
\end{split}
\label{eq: para S matrix}
\end{equation}
Here, the four matrices $\sigma^{\ }_{0,1,2,3}$ 
act on the spin-$1/2$ components with
$\sigma^{\ }_{0}$ the unit $2\times2$ matrix 
and $(\sigma^{\ }_{1},\sigma^{\ }_{2},\sigma^{\ }_{3})$ 
the three Pauli matrices.
Moreover,
$0\leq{X}<\infty$,
$0\leq\theta<\pi/2$,
$0\leq\varphi^{\ }_{1}<2\pi$,
and
$0\leq\varphi^{\ }_{2}<2\pi$.
TRS is represented by the condition
\begin{equation}
S
=
\left(
\begin{array}{cc}
\sigma^{\ }_{2}
& 
0
\\
0
& 
\sigma^{\ }_{2}
\end{array}
\right)
S^{T}
\left(
\begin{array}{cc}
\sigma^{\ }_{2}
& 
0
\\
0
& 
\sigma^{\ }_{2}
\end{array}
\right).
\end{equation}
The matrix $S$ is the most general $4\times4$ unitary matrix
that describes a quantum tunneling process between
two Kramers doublets that preserves TRS but can break SRS.
When $r=1$ and $t=0$, $S$ is reduced to the unit matrix,
and there is no tunneling between one Kramers
pair $(\psi^{\ }_{1}+\psi^{\ }_{2})$ and
the other pair $(\psi^{\ }_{3}+\psi^{\ }_{4})$.
The tunneling with (without) a spin flip occurs with the
probability $t^2\sin^2\theta$ ($t^2\cos^2\theta$).
Although $S$ is parametrized by four real
parameters, only $X$ and $\theta$ matter
as $\varphi^{\ }_{1}$ and $\varphi^{\ }_{2}$ can
be absorbed in the overall phase windings that
the Kramers doublets acquire when traversing along
links connecting nodes.

Disorder is introduced in the network model by assuming
that the phases of Kramers doublets on the links
and the phases $\theta$
on the nodes are independently and identically distributed.
The distribution of the link phase of a Kramers doublet is
uniform over the interval $[0,2\pi[$. The distribution of
$\theta$ is $\sin2\theta$ over the interval $[0,\pi/2]$.
We are left with one parameter $X$
in the network model that controls the scattering
amplitude at every nodes.
The parameter $\theta$ in the network model
plays the same role as
spin-orbit interactions (of Rashba type)
in a random tight-binding Hamiltonian belonging 
to the two-dimensional symplectic symmetry class,
whereas the parameter $X$
plays the role of the Fermi energy.

In order to distinguish the topologically-trivial insulating phase
from the QSH insulating phase,
we shall compare the results that we obtained from the network model
against the ones that we obtained from a two-dimensional
tight-binding model introduced in Ref.~\onlinecite{Asada02}, 
the so-called SU(2) model. The SU(2) model 
is a microscopic random tight-binding Hamiltonian
with on-site randomness (box distributed with the width $W$)
and with random hopping amplitudes 
such that the spin-dependent hopping amplitudes between any
pair of nearest-neighbor sites of a square lattice 
are distributed so as to generate the SU(2) invariant Haar measure.
It captures the transition between a metallic 
and a topologically-trivial insulating state 
in the ordinary two-dimensional symplectic universality class.
In the SU(2) model, the Fermi energy plays the role of 
the network parameter $X$ for a fixed and not too strong $W$.

\section{
Phase diagram and boundary conditions
        }
\label{sec: Phase diagram}

\begin{figure}[t]
\centering
\includegraphics[width=0.45\textwidth]{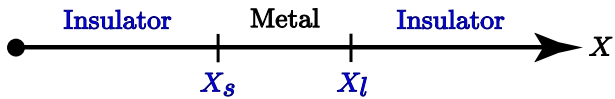}
\caption{
(Color online)
Phase diagram for the network model after 
Ref.~\onlinecite{Obuse07b}.
        } 
\label{fig:bulk phase diagram}
\end{figure}

\begin{figure}[h]
\centering
\includegraphics[width=0.45\textwidth]{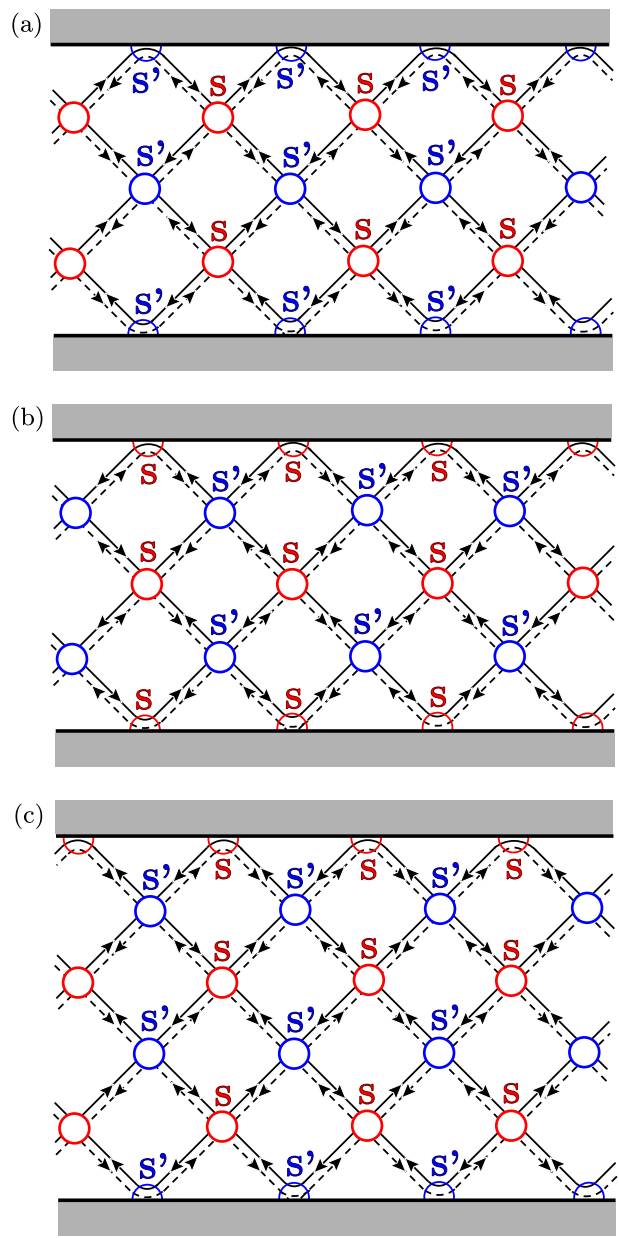}
\caption{
(Color online)
Quasi-one-dimensional network model with reflecting boundaries.
(a) The upper and lower boundaries pass through the nodes $\textsf{S}'$. 
(b) The upper and lower boundaries pass through the nodes $\textsf{S}$.
(c) The upper and lower boundaries pass through the nodes $\textsf{S}$
and $\textsf{S}'$, respectively. 
Transverse periodic boundary conditions are only possible in
geometries (a) or (b) for which the transverse width is here $M=2$.
        }
\label{fig:3Q1D geo}
\end{figure}

On symmetry grounds, we expect that 
it is possible to drive the network model through
two successive Anderson transitions by tuning $X$.
Indeed, this was shown to be the case in Ref.%
~\onlinecite{Obuse07b}.
For any $X$ bounded by the two critical values 
$X^{\ }_{s}<X^{\ }_{l}$
[$X^{\ }_{s}=0.047\pm0.001$, $X^{\ }_{l}=0.971\pm0.001$;
the subscript $s$ ($l$) stands for small (large)]
the network model is in a metallic state,
while for $X<X^{\ }_{s}$ or for $X^{\ }_{l}<X$
it is in an insulating state. It was shown in Ref.%
~\onlinecite{Obuse07b}
that the localization length, a bulk property of 
any one of the two insulating phases,
diverges with the exponent $\nu\approx2.7$
of the ordinary two-dimensional symplectic universality
class\cite{Asada02} upon approaching the mobility edge 
(either $X^{\ }_{s}$ or $X^{\ }_{l}$).
The resulting phase diagram is shown in 
Fig.~\ref{fig:bulk phase diagram}.
The conclusions from Ref.%
~\onlinecite{Obuse07b}
were reached by studying numerically the
network model in a strip geometry
with periodic boundary conditions (PBC)
imposed in the transverse direction.

\begin{figure}[t]
\centering
\includegraphics[width=0.45\textwidth]{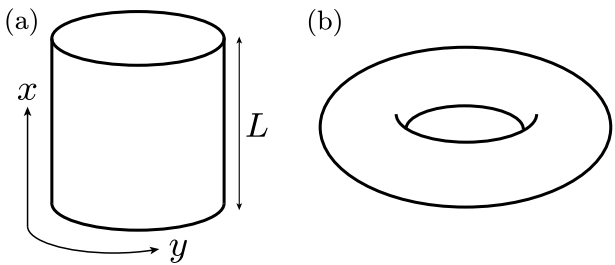}
\caption{
(a)
The network model on a square with $L^{2}$ nodes of type $\textsf{S}$
and $L^2$ nodes of type $\textsf{S}'$
(a node on a boundary is counted as 1/2)
is wrapped around a cylinder by imposing
reflecting boundary conditions (RBC)
along the ${x}$ direction and periodic boundary conditions (PBC)
along the ${y}$ direction.
(b) 
It is wrapped around a torus
by imposing PBC in the ${x}$ and ${y}$ directions.
        } 
\label{fig:cyl-torus}
\end{figure}

Here, we shall depart from Ref.~\onlinecite{Obuse07b}
by imposing reflecting boundary conditions (RBC)
in the transverse direction in the strip geometry
such as in Fig.~\ref{fig:3Q1D geo}
or Fig.~\ref{fig:cyl-torus}(a) for example.
Reflecting boundaries for the network model 
that respect the TRS are defined by choosing a set of nodes
and amputating two pairs of links
(one labeled by an odd integer and the other by an even integer
in the convention of Fig.~\ref{fig:nodeS})
attached to any of these nodes 
in a consistent fashion, 
i.e., no node can be the end point of an odd number of pairs of links. 
A boundary node is depicted by a colored semi-circle,
and its scattering matrix $S$ is always represented by
a unit $2\times2$ matrix, regardless of its index $\textsf{S}$ or
$\textsf{S}'$.
Three examples of reflecting boundaries
defining a strip geometry
are shown in Fig.~\ref{fig:3Q1D geo}. From the fact that
the scattering process in Fig.~\ref{fig:nodeS}(b)
is obtained by a clockwise rotation by $\pi/2$ of 
the scattering process in Fig.~\ref{fig:nodeS}(a)
follows the important property that a horizontal boundary passing
through nodes of type $\textsf{S}$ is equivalent to a vertical boundary
passing through the nodes of type $\textsf{S}'$ and vice versa.

In the following discussions we shall focus on the critical
point $X=X^{\ }_{l}$ and its neighboring insulating phase.
Imposing RBC,
we shall see that the insulating phase at $X>X^{\ }_{l}$ in
the phase diagram from Fig.~\ref{fig:bulk phase diagram}
can acquire a topological attribute in that it supports 
a single Kramers doublet that is localized in the direction transverse
to the boundary but is delocalized along the boundary, i.e.,
a single Kramers degenerate pair of edge states.

\begin{figure}[h]
\centering
\includegraphics[width=0.45\textwidth]{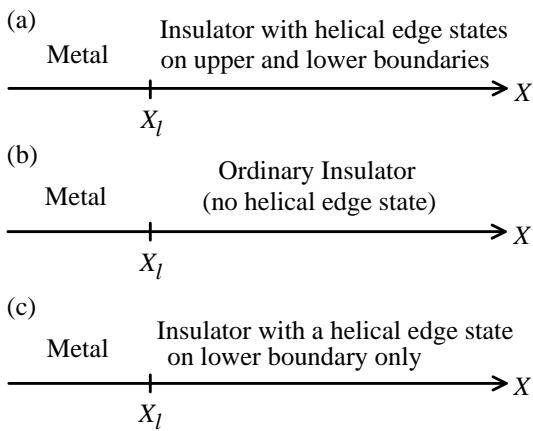}
\caption{
Phase diagram of the network model of the
quasi-one-dimensional geometries 
in Fig.~\ref{fig:3Q1D geo}(a),
\ref{fig:3Q1D geo}(b),
and
\ref{fig:3Q1D geo}(c),
respectively,
at $X\gtrsim X^{\ }_{l}$.
The choice of the boundary decides whether the insulating phase
at $X>X^{\ }_{l}$ has helical edge states.
        } 
\label{fig:refined phase diagram}
\end{figure}

\begin{figure}[t]
\centering
\includegraphics[width=0.45\textwidth]{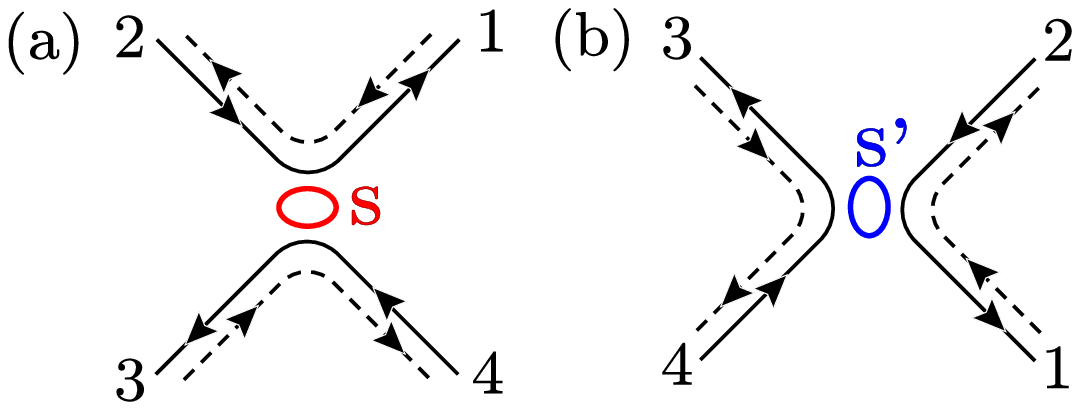}
\caption{
(Color online)
Scattering in the reflective limit $X\to\infty$
for node $\textsf{S}$ (a) and $\textsf{S}'$ (b).
        } 
\label{fig:nodeS if X large}
\end{figure}

\begin{figure}[t]
\centering
\includegraphics[width=0.45\textwidth]{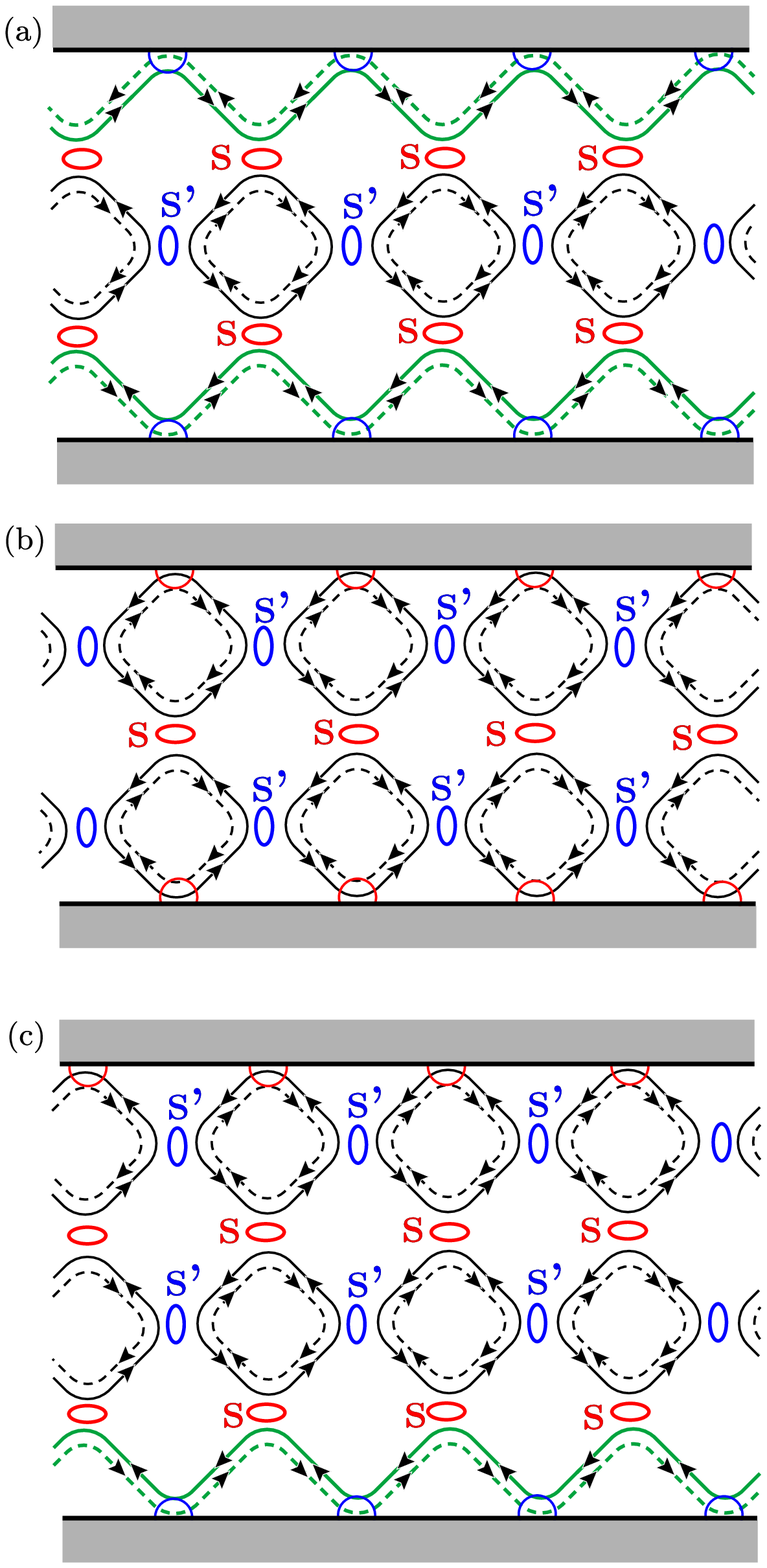}
\caption{
(Color online)
Large $X$ limit of the
quasi-one-dimensional network model with the reflecting
boundaries from Fig.~\ref{fig:3Q1D geo}.
        } 
\label{fig:3Q1D geo if X large}
\end{figure}

The refined phase diagram Fig.~\ref{fig:refined phase diagram}
is depicted for $X\gtrsim X^{\ }_{l}$
according to the presence or absence of edge states
in the insulating phase at $X>X^{\ }_{l}$.
Identifying {helical edge states can be done
in a pictorial way 
in the perfect reflection limit $X\to\infty$,
provided we assume continuity of the insulating phase.
In the limit $X\to\infty$, 
the scattering at the nodes  $\textsf{S}$ and $\textsf{S}'$
reduces to the glancing events depicted in
Fig.~\ref{fig:nodeS if X large}.
As a corollary
Fig.~\ref{fig:3Q1D geo}
simplifies to Fig.~\ref{fig:3Q1D geo if X large}.
We thus see two Kramers doublets propagating along
the upper and lower boundaries
in Fig.~\ref{fig:3Q1D geo if X large}(a),
none in Fig.~\ref{fig:3Q1D geo if X large}(b),
and one in Fig.~\ref{fig:3Q1D geo if X large}(c).
The refined phase diagrams from Fig.~\ref{fig:refined phase diagram}
then follow assuming continuity.\cite{note on the opposite limit}
In the following sections we go beyond the pictorial argument from
Fig.~\ref{fig:3Q1D geo if X large} and
study critical properties at the Anderson metal-insulator transition
in a quantitative fashion
with an emphasis on the boundary and corner multifractal
scaling exponents at criticality.

It is important to note here that we may consider the insulating
state in Fig.~\ref{fig:refined phase diagram}(a) as a $\mathbb{Z}^{\ }_{2}$
topological insulating state because of the presence of a helical
edge state along the boundaries.
Similarly, we may identify the insulating state in
Fig.~\ref{fig:refined phase diagram}(b) with a $\mathbb{Z}^{\ }_{2}$
topologically trivial, ordinary insulator.
These observations indicate that, in the network model formulation,
the $\mathbb{Z}^{\ }_{2}$ topological nature of an insulating state can be
determined by appropriate choice of reflecting boundary nodes,
even without changing the control parameter $X$.
This is very similar to the situation in the
Chalker-Coddington network model\cite{Chalker88} 
for an integer quantum Hall plateau transition,
in which an insulating state may or may not have an edge state
for a given value of quantum tunneling parameter at nodes,
depending on the location of reflecting boundary nodes;
an insulating state with an edge state can be regarded
as an integer quantum Hall state.
We also note that we may attribute this
dependence of the topological nature of an insulator on the
reflecting boundary conditions in network models to our freedom
to give any topological character to the vacuum outside
the insulator.

\section{
Normalized localization length
        }
\label{sec: Localization length}

Supporting evidences for the phase diagram in 
Fig.~\ref{fig:refined phase diagram}(a)
can be extracted from the dependence on $M$ of
\begin{equation}
\Lambda^{(i)}(X,M)\equiv 
\xi^{(i)}(X,M)/M
\label{eq: def Lambda(i)}
\end{equation}
in the geometry of 
Fig.~\ref{fig:3Q1D geo}(a).
In Eq.~(\ref{eq: def Lambda(i)}),
the $i$th normalized localization length $\Lambda^{(i)}(X,M)$
is the ratio between
the $i$th localization length $\xi^{(i)}(X,M)$, which is
given by the inverse of the value of 
the $i$th smallest pair\cite{footnote:Kramers degeneracy Lyapunov} 
of Lyapunov exponent of the transfer matrix,
and the width $M$ in either one of the 
geometries of Figs.~\ref{fig:3Q1D geo}(a) or \ref{fig:3Q1D geo}(b).%
~\cite{MacKinnon83}
The transfer matrix for the strip geometries can be constructed
in a similar manner as in the cylinder geometry (transverse PBC).%
~\cite{Obuse07b}

\begin{figure}[t]
\centering
\includegraphics[width=0.45\textwidth]{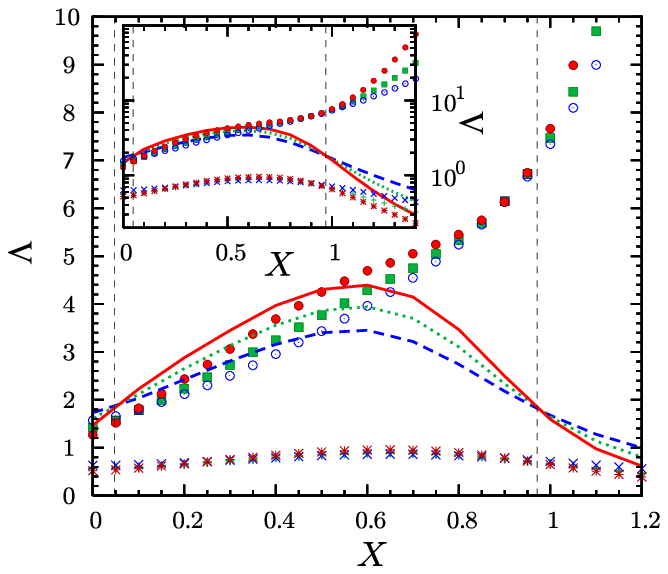}
\caption{
(Color online)
The dependence on $X$ of $\Lambda^{(1)}(X,M)$ in the geometry
of Fig.~\ref{fig:3Q1D geo}(a) 
with transverse RBC is shown for 
$M=4$ ($\textcolor{blue}{\odot}$), 
$8$ ($\textcolor[rgb]{0,0.7,0.25}{\blacksquare}$), 
and $16$ (\textcolor{red}{$\bullet$}).
The dependence on $X$ of $\Lambda^{(2)}(X,M)$ 
with transverse RBC is shown for  
$M=4$ ($\textcolor{blue}{\times}$), 
$8$ ($\textcolor[rgb]{0,0.7,0.25}{+}$), 
and $16$ ($\textcolor{red}{*}$).
The dependence on $X$ of $\Lambda^{(1)}(X,M)$ 
with transverse PBC 
is shown for 
$M=4$ (blue dashed curve), 
$8$ (green dotted curve), 
and $16$ (red solid curve).
The vertical dashed lines identify the critical points 
$X^{\ }_{s}$ and $X^{\ }_{l}$ 
deduced in Ref.~\onlinecite{Obuse07b}
when transverse PBC are imposed.
The inset displays the same data points with a logarithmic vertical scale.
        } 
\label{fig:Lambda}
\end{figure}

Figure~\ref{fig:Lambda} shows the dependence on $X$ of
$\Lambda^{(i)}(X,M)$ with $i=1$ and $i=2$
for the values of $M=4$, $8$, and $16$, respectively,
in the geometry of Fig.~\ref{fig:3Q1D geo}(a).
Also shown in Fig.~\ref{fig:Lambda} is the dependence of
$\Lambda^{(1)}(X,M)$ for transverse PBC.
The existence of the critical points $X^{\ }_{s}$ and  $X^{\ }_{l}$
is signaled by the independence of 
$\Lambda^{(1)}(X^{\ }_{s,l},M)$
on $M$. Imposing transverse PBC reduces the finite-size corrections
and allows a more accurate measurement of
$X^{\ }_{s}$ and  $X^{\ }_{l}$, the dashed vertical lines
in Fig.~\ref{fig:Lambda}.%
~\cite{Obuse07b}
When $X^{\ }_{s}<X<X^{\ }_{l}$,
$\Lambda^{(i)}(X,M)$ increases with $M$ 
for both transverse RBC and transverse PBC,
as is expected from
a two-dimensional metallic state.
This trend is reversed in the insulating phases
for transverse PBC, as is expected from
a two-dimensional insulating state.
However, when $X^{\ }_{l}<X$ and transverse RBC
are chosen, 
$\Lambda^{(1)}(X,M)$
remains an increasing function of $M$
whereas 
$\Lambda^{(2)}(X,M)$
becomes a decreasing function of $M$
(the inset shows this more clearly with its logarithmic vertical scale).
This opposite dependence on $M$ of $\Lambda^{(1)}(X,M)$ and
$\Lambda^{(2)}(X,M)$ 
is the signal that each of the upper and lower boundaries
supports a \textit{single} Kramers doublet that would
be extended along the boundary,\cite{Zirnbauer92,Takane04} 
were it not for the existence of a finite tunneling
amplitude (i) that couples
the two Kramers doublets (edge states)
residing near the two opposite boundaries
and (ii)
that is exponentially small in $M$ when $M$ 
is much longer than the mean free path.
The dependence on $M$ of  
$\Lambda^{(1)}(X,M)$
with transverse RBC when $X<X^{\ }_{s}$
is the one expected from an ordinary insulating state.
All together, these observations point to the 
refined phase diagram
shown in Fig.~\ref{fig:refined phase diagram}(a), i.e.,
the insulating phase is unconventional when $X^{\ }_{l}<X$
due to the presence of a single Kramers doublet of
edge states per boundary that becomes delocalized along 
the boundaries in the limit $M\to\infty$.

The $M$-independent,
normalized localization length at the Anderson transition
is nothing but the normalized correlation length $\Lambda^{\ }_{c}$
at a generic continuous phase transition in the theory of critical phenomena. 
This quantity is known to depend on the choice of the 
transverse boundary conditions.\cite{Cardy84}
First, the value 
\begin{equation}
\Lambda^{\mathrm{(pbc)} }(W^{\ }_{c})=
1.844\pm0.002
\label{eq: su(2) pbc Lambda}
\end{equation}
of the normalized localization length at the Anderson transition
was calculated in Ref.~\onlinecite{Asada02}
for the SU(2) model with transverse PBC. 
It should here be compared with the values
\begin{equation}
\begin{split}
&
\Lambda^{\mathrm{(pbc)}}(X^{\ }_{s})=
1.81\pm0.01,
\\
&
\Lambda^{\mathrm{(pbc)}}(X^{\ }_{l})=
1.82\pm0.01,
\end{split}
\label{eq: net pbc Lambda}
\end{equation}
calculated in Ref.~\onlinecite{Obuse07b}.
Second,
we have extended the calculation in Ref.~\onlinecite{Asada02}
for the SU(2) model in which transverse PBC were used to the case
in which the vanishing of the wave functions along the
transverse boundaries is imposed, a situation that we shall refer
to as fixed boundary conditions (FBC),
and found
[see Fig.~\ref{fig: Lambdac OBC}(c)]
\begin{equation}
\Lambda^{\mathrm{(fbc)}}(W^{\ }_{c})=
1.50\pm0.03.
\label{eq: su(2) obc Lambda} 
\end{equation}
This value should be compared with
\begin{equation}
\Lambda^{\mathrm{(rbc)}}(X^{\ }_{s})=
1.49\pm0.02
\label{eq: net obc Lambda} 
\end{equation}
for the network model with transverse RBC
[where we can also
deduce the right-hand side 
from Fig.~\ref{fig: Lambdac OBC}(c) for which
it is the network model at the critical point $X^{\ }_{l}$
in the geometry of Fig.~\ref{fig:3Q1D geo}(b)
that is investigated; 
Fig.~\ref{fig: Lambdac OBC}(a) shows insulating behavior
of $\Lambda^{(1)}$ for $X>X^{\ }_{l}$].
Evidently, the values~(\ref{eq: su(2) obc Lambda})
and~(\ref{eq: net obc Lambda})
agree within their error bars
(in support of our identification,
made at the end of the last section,
of the insulating phase without a helical edge state
with a $\mathbb{Z}^{\ }_{2}$ topologically trivial insulator)
but clearly differ from the
values~(\ref{eq: su(2) pbc Lambda})
and (\ref{eq: net pbc Lambda}).
Finally, and more importantly, there is a rather large asymmetry 
\begin{equation}
\Lambda^{\mathrm{(rbc)}}(X^{\ }_{l})/
\Lambda^{\mathrm{(rbc)}}(X^{\ }_{s})
\approx
4.8
\label{eq:asymmetric Lambda}
\end{equation}
for the network model with transverse RBC
in the geometry of
Fig.~\ref{fig:3Q1D geo}(a).

\begin{figure}[t]
\centering
\includegraphics[width=0.45\textwidth]{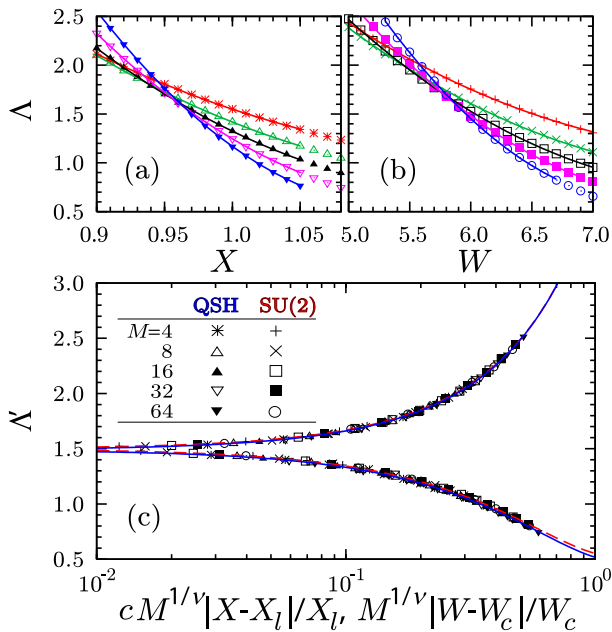}
\caption{
(Color online)
(a)
Dependence on $X$ of the normalized localization length 
$\Lambda^{(1)}(X,M)\equiv \Lambda(X,M)$
for the network model in the quasi-one-dimensional geometry  
of Fig.~\ref{fig:3Q1D geo}(b) 
with transverse RBC and $M=4,8,16,32,64$.
(b) 
Dependence on the dimensionless disorder strength $W$ 
of the normalized localization length 
$\Lambda^{(1)}(W,M)\equiv\Lambda(W,M)$
for the SU(2) model in the
quasi-one-dimensional geometry 
with transverse fixed boundary conditions (FBC)
and $M=4,8,16,32,64$.
The solid curves are computed from the finite-size scaling functions
that follow.
(c)
Finite-size scaling analysis for the data shown in (a-b):
The solid blue curves represent the scaling functions for 
the network model with RBC
on node $\textsf{S}$ in the vicinity of the
critical point $X^{\ }_{l}$ where we find
$\Lambda^{(\mathrm{rbc})}(X^{\ }_{l})=1.49\pm0.02$ 
and 
$\nu=2.88\pm0.04$.
The red dashed curves represent the 
scaling functions for the SU(2) model with FBC
in the vicinity of the critical point $W^{\ }_{c}$
where we find
$\Lambda^{(\mathrm{rbc})}(W^{\ }_{c})=1.50\pm0.03$ and $\nu=2.85\pm0.06$.
The scaling functions are obtained from finite-size scaling analysis
incorporating corrections from a leading irrelevant scaling variable.%
~\cite{Slevin,Obuse07b}
The normalized localization length
$\Lambda'$ is obtained from $\Lambda$ by subtracting 
these corrections,
as defined in Eqs.\ (3.7) and (3.8) from Ref.~\onlinecite{Obuse07b}.
The distance to the critical point $|X-X^{\ }_l|$ is rescaled by
a factor $c\simeq 1.7$
in the scaling function for the network model with RBC
so that it coincides with that for the SU(2) model with FBC.
        } 
\label{fig: Lambdac OBC}
\end{figure}

Finite-size scaling in Fig.~\ref{fig: Lambdac OBC}(c) clearly shows that
the scaling function obtained for the SU(2) model with transverse
FBC is identical to that for the network model at $X=X^{\ }_{l}$
in the geometry of Fig.~\ref{fig:3Q1D geo}(b).
These scaling functions are therefore universal property
of the critical point between a metallic phase and an ordinary insulator.
From this finite-scaling analysis we also obtained
the critical exponent $\nu$ of the diverging localization length.
At the critical point $X^{\ }_{l}$ of the network model we found 
$
\nu=2.88\pm0.04.
$
Again, this should be compared with
the critical exponent $\nu$ of the SU(2) model with transverse
FBC for which we find
$
\nu=2.85\pm0.06
$.
Both exponents agree with each other within their error bars;
they are also consistent with the exponent $2.7\lesssim\nu\lesssim2.8$
obtained with transverse PBC.\cite{Asada02,Obuse07b}
This implies that the exponent $\nu$ is a bulk property
independent of boundary conditions
(whereas the scaling functions are dependent on the
transverse boundary conditions).

The network and SU(2) models
in a quasi-one-dimensional cylinder geometry
at criticality are indistinguishable
as measured by the normalized localization length%
~(\ref{eq: su(2) pbc Lambda}) and (\ref{eq: net pbc Lambda}),
respectively. The same is true in a strip geometry,
provided the insulating side of the transition in the network
model is topologically trivial
(i.e., has no helical edge states)
according to Eqs.%
~(\ref{eq: su(2) obc Lambda})
and (\ref{eq: net obc Lambda}).
However, the strong asymmetry~(\ref{eq:asymmetric Lambda})
hints at the possibility that some boundary critical exponents might
be sensitive to the choice of transverse boundary conditions 
that dictates the presence or absence 
of a delocalized Kramers doublet of edge states.

\section{
Bulk and boundary criticality
        }
\label{sec: Bulk and boundary criticality}

\subsection{
Typical spatial profile of wave functions 
           }

To gain more insights on the criticality of the network model,
we study numerically the critical normalized wave functions
$\Psi$ when $X=X^{\ }_{l}$ 
in the geometries of Figs.~\ref{fig:cyl-torus}(a)
or \ref{fig:cyl-torus}(b).
To this end, the support of normalized wave functions for the network model
is defined on the midpoints $({x},{y})$
of every bond joining nearest-neighbor nodes,
i.e., the normalized wave function
$\Psi$ can be viewed as a complex-valued vector whose $8L^2$ components 
$\psi^{\ }_{\sigma}({x},{y})$,
with the spin-1/2 quantum number labeled by 
$\sigma=\uparrow,\downarrow$, 
correspond to $4L^2$ freely propagating Kramers pairs of plane waves.
The dynamics of such a wave function is governed by a
unitary evolution operator $U$ built out of all scattering matrices at
the nodes of the network.\cite{Klesse95}
For each realization of the disorder, 
we numerically diagonalize $U$ and retain
one normalized wave function whose eigenvalue is closest to $1$.
The number of disorder realizations is $5\times 10^4$ 
for each system size $L$ ($L$ ranging from 20 to 80).
By tuning the parameter  
$X=X^{\ }_{l}$ at each node, this normalized wave function is, 
with a slight abuse of language, called critical.

\begin{figure}[t]
\centering
\includegraphics[width=0.45\textwidth]{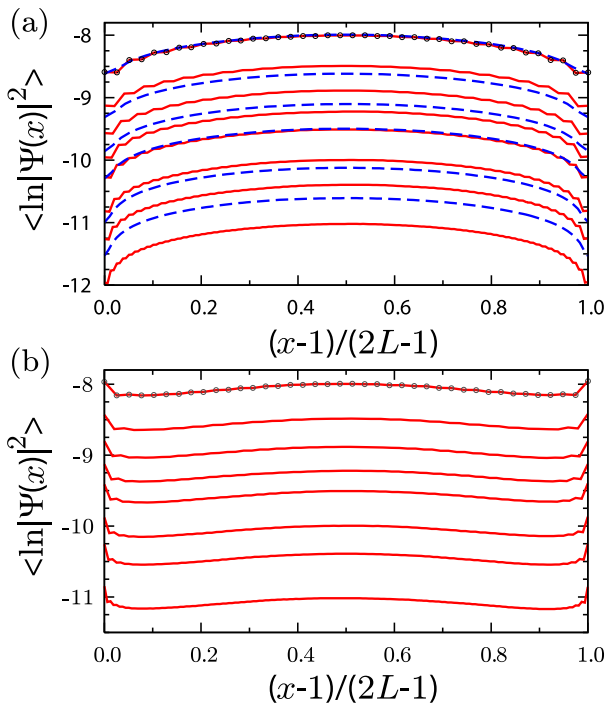}
\caption{
(Color online)
Dependence on the position ${x}$ 
along the cylinder axis of Fig.~\ref{fig:cyl-torus}(a)
of 
$
\left\langle
\ln|\Psi|^2
\right\rangle^{\ }_{{x},L}
$
defined by Eq.~(\ref{eq: def <ln|Psi|2>})
for the values of
$L=20,25,30,35,40,50,60,80$ from top to bottom with
(a) $\textsf{S}$ boundaries
or with (b) $\textsf{S}'$ boundaries.
The blue dashed curves in (a)
represent the normalized critical wave functions 
$
\left\langle
\ln|\Psi|^2
\right\rangle^{\ }_{{x},L}
$
for $L=36,48,60,72,96,120$
obtained from the SU(2) model defined in Ref.~\onlinecite{Asada02}
using the cylinder geometry of Fig.~\ref{fig:cyl-torus}(a).
       } 
\label{fig:WFA}
\end{figure}

First, we define
\begin{equation}
\left\langle
\ln|\Psi|^2
\right\rangle^{\ }_{{x},L}\equiv 
(2L)^{-1} 
\sum_{{y}=1}^{2L} 
\overline{
\ln
\Bigg(
\sum_{\sigma=\uparrow,\downarrow}|\psi^{\ }_{\sigma}({x},{y})|^2
\Bigg)
         },
\label{eq: def <ln|Psi|2>}
\end{equation}
where $X$ is tuned to the critical point $X^{\ }_{l}$
for the network model.
The overline denotes disorder averaging over the
nodes through the independently distributed angles $\theta$
as well as over the links through the independently 
and uniformly distributed random phases of Kramers doublets.

Figure~\ref{fig:WFA} shows the ${x}$ dependence of
$
\left\langle
\ln|\Psi|^2
\right\rangle^{\ }_{{x},L}
$
for different values of $L$
at the critical point $X^{\ }_{l}$
in the cylinder geometry of Fig.~\ref{fig:cyl-torus}(a).
The two figures correspond to different RBC
in the $x$ direction:
(a) the boundaries passing through the \textsf{S} nodes
as in Fig.~\ref{fig:3Q1D geo}(b), and
(b) the boundaries passing through the $\textsf{S}'$ nodes
as in Fig.~\ref{fig:3Q1D geo}(a).
The boundary bonds are located at ${x}=1$ and ${x}=2L$.
At the critical point between the metallic
and the insulating state without helical edge state
[Fig.~\ref{fig:WFA}(a)],
the dependence of 
$
\left\langle
\ln|\Psi|^2
\right\rangle^{\ }_{{x},L}
$
on ${x}$ is symmetric about ${x}=L$ 
where it reaches a maximum, 
and reaches a minimum close to the boundaries.
For comparison, we show in Fig.~\ref{fig:WFA}(a)
$\left\langle\ln|\Psi|^2\right\rangle^{\ }_{x,L}$ of the
SU(2) model, in which the on-site disorder strength $W$ is tuned 
to the critical point $W^{\ }_{c}$, and
the overline in Eq.\ (\ref{eq: def <ln|Psi|2>}) denotes
disorder averaging over the on-site energies and the Haar measure
of the hopping matrix elements.
We find no qualitative difference in the dependence 
on $x$ and $L$ of
$
\left\langle
\ln|\Psi|^2
\right\rangle^{\ }_{x,L}
$
between the two models when the critical point
separates a metal and an ordinary insulator.
Remarkably, 
Fig.~\ref{fig:WFA}(b)
shows that the dependence on $x$ 
of
$
\left\langle
\ln|\Psi|^2
\right\rangle^{\ }_{{x},L}
$
is nonmonotonic when approaching
a boundary from the center of the bulk $x=L$
and that its absolute maximum is reached at the boundaries ${x}=1,2L$
instead of the local maximum at ${x}=L$.

\subsection{
Scaling exponents $\Delta^{\ }_{q}$
           }

This striking difference between Fig.~\ref{fig:WFA}(a) and
Fig.~\ref{fig:WFA}(b)
suggests that the statistics of the critical normalized wave functions
is sensitive to the topological nature
(i.e., the presence or absence of helical edge states)
of the insulating state.
A powerful tool to study the statistics of critical normalized wave functions
is the multifractal scaling analysis.%
~\cite{Janssen94}
Multifractality is encoded by a set of 
scaling exponents (anomalous dimensions)
$\Delta^{(\kappa)}_{q}$
defined by the scaling laws
\begin{equation}
\overline{
\left|\Psi(\bm{r})\right|^{2q}
         } 
\Big/ 
{\left(\overline{\left|\Psi(\bm{r})\right|^2}\right)}^q
\propto
L^{-\Delta^{(\kappa)}_{q}}
\label{eq: def Deltaq}
\end{equation}
for $L\gg1$,
where $\Psi$ is the normalized spinor obtained from the network model
in the geometries~\ref{fig:cyl-torus}(a) or \ref{fig:cyl-torus}(b)
at criticality $X=X^{}_{l}$,
and $q$ is any real number.
The index $(\kappa)$ distinguishes
bulk, boundary, and corner exponents
through the choice of the location of
$\bm{r}=({x},{y})$ 
relative to the boundaries and the nature of the boundaries.%
~\cite{Subramaniam06,Mildenberger07b,Obuse07a,Evers08,Obuse08}
In this paper, we distinguish six cases, 
$(\kappa)=(2)$,
$(1,\mathbb{Z}^{\ }_{2})$,
$(1,\mathrm{O})$,
$(0,\mathbb{Z}^{\ }_{2})$,
$(0,\mathrm{O})$,
and
$(0,\mathbb{Z}^{\ }_{2}|\mathrm{O})$.
Here the first entry in $(\kappa)$
refers to the dimensionality of the support of
wave functions.
The second entry in $(\kappa)$,
$\mathbb{Z}^{\ }_{2}$ or O (``ordinary''),
refers to the presence or absence, on the insulating
side of the critical point, of a Kramers doublet (helical) edge state
along a boundary or at a corner where the measurement (\ref{eq: def Deltaq})
is made (see below for more detailed definitions).

In Eq.~(\ref{eq: def Deltaq})
the proportionality constant may have
a dependence on $L$ that is much milder
than the power law it multiplies.
This means that the scaling exponents
$\Delta^{(\kappa)}_{q}$ are obtained from calculating numerically
\begin{equation}
D^{(\kappa)}_{q}(L)\equiv
\frac{
q\,
\ln
\overline{
\overline{
\left|\Psi(\bm{r})\right|^{2}
         } 
         } 
-
\ln
\overline{
\overline{
\left|\Psi(\bm{r})\right|^{2q}
         } 
         } 
     }
     {
\ln L
     }
\end{equation}
first. Here, the double overline means that in addition to the disorder
averaging a spatial average over the relevant sites $\bm{r}$
($4L^{2}$ sites for the torus geometry, $2L$ sites for the boundary)
is also taken to improve the statistics.
This is followed by a linear fit of
$D^{(\kappa)}_{q}(L)$
as a function of $1/\ln L$.
Only system sizes larger than 
$L=35$ are kept for the linear fit.
At last, $\Delta^{(\kappa)}_{q}$ is obtained as
the intercept of the linear fit with the vertical axis at 
$1/\ln L=0$. For the case of boundary scaling exponents,
we also replaced $|\Psi(\bm{r})|^{2}$ 
with $\bm{r}$ a bond joining a node on the boundary by 
averaging it over all bonds from the elementary plaquette 
to which it belongs. This coarse graining is necessary to 
overcome the small oscillations that are visible in the ${x}$
dependence plotted in Fig.~\ref{fig:WFA}
close to the boundaries.

\begin{figure}[t]
\centering
\includegraphics[width=0.45\textwidth]{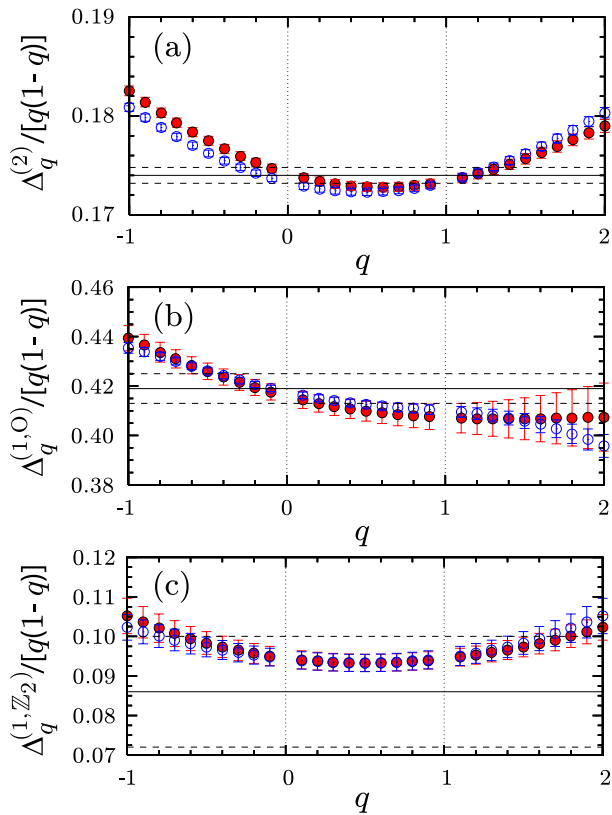}
\caption{
(Color online)
The dependence on $q$ of $\Delta^{(\kappa)}_{q}/[q(1-q)]$ 
($\textcolor{red}{\bullet}$) 
when $X=X^{\ }_{l}$ for 
(a) the bulk: $(\kappa)=(2)$, 
(b) the $\textsf{S}$ boundary: $(\kappa)=(1,\mathrm{O})$, 
and (c) the $\textsf{S}'$ boundary:
$(\kappa)=(1,\mathbb{Z}^{\ }_{2})$.
The solid lines represent $\alpha^{(\kappa)}(q)-2$
evaluated at $q=0$ according to Eq.~(\ref{eq:def alpha})
with its error bars indicated by dashed lines.
The dependence on $q$ of $\Delta^{(\kappa)}_{q}/[q(1-q)]$ 
for the SU(2) model in the bulk and on the boundary
are shown with $\textcolor{blue}{\circ}$
in (a) and (b), respectively.
Blue open circles $\textcolor{blue}{\circ}$
in (c) show the $q$ dependence of 
$\Delta^{(1,\mathbb{Z}^{\ }_{2})}_{1-q}/[q(1-q)]$. 
        } 
\label{fig:Delta}
\end{figure}

Figure~\ref{fig:Delta}
shows the $q$ dependence of the scaling exponents 
$\Delta^{(\kappa)}_{q}$
for the network and the SU(2) model
in different geometries.

The bulk scaling exponents,
for which case $(\kappa)=(2)$,
characterize the scaling law (\ref{eq: def Deltaq})
in the torus geometry from Fig.~\ref{fig:cyl-torus}(b)
(which has no boundary).
This is also the case in the cylinder geometry of
Fig.~\ref{fig:cyl-torus}(a), if the distance from the site $\bm{r}$
to the boundaries is of order $L$.
The bulk exponents shown in Fig.~\ref{fig:Delta}(a)
are obtained with the torus geometry
in order to be free from boundary effects
and to maximize the number of sampling points $\bm{r}$.
Figure~\ref{fig:Delta}(a)
shows that the bulk scaling exponents $\Delta^{(2)}_q$
of the network model are identical to those obtained for
the SU(2) model\cite{Obuse07a}
within their error bars. 

The boundary scaling exponents are obtained
by putting $\bm{r}$
on the boundaries (with the coarse graining mentioned above)
in the cylinder geometry from
Fig.~\ref{fig:cyl-torus}(a).
There is an important caveat however.
As we saw in the preceding sections,
the network model supports two types of (straight) boundaries,
while the SU(2) model has only one.
Indeed, for the network model
at $X=X^{\ }_{l}$ studied numerically
in Fig.~\ref{fig:Delta} we have two cases:
\begin{enumerate}
\item[(i)]
$(\kappa)=(1,\mathrm{O})$, when
the boundary on which $\bm{r}$ is located
passes through the nodes of type $\textsf{S}$.
To reduce statistical errors, we
take the cylinder geometry of Fig.~\ref{fig:cyl-torus}(a)
which follows from imposing longitudinal PBC
in Fig.~\ref{fig:3Q1D geo}(b).
No helical edge state exists on the boundaries
in the insulating phase at $X>X^{\ }_{l}$.
\item[(ii)]
$(\kappa)=(1,\mathbb{Z}^{\ }_{2})$, when
the boundary on which $\bm{r}$ is located
passes through the nodes of type $\textsf{S}'$.
To reduce statistical errors, we
take the cylinder geometry of Fig.~\ref{fig:cyl-torus}(a)
which follows from imposing longitudinal PBC
in Fig.~\ref{fig:3Q1D geo}(a).
There exists a helical edge mode on each boundary
in the insulating phase at $X>X^{\ }_{l}$.
\end{enumerate}
Figure~\ref{fig:Delta}(b)
shows that the boundary scaling exponents $\Delta^{(1,\mathrm{O})}_q$
in the network model
agree, within their error bars,
with the boundary scaling exponents obtained
in Ref.~\onlinecite{Obuse07a} from
the SU(2) model in the cylinder geometry of 
Fig.~\ref{fig:cyl-torus}(a).
Remarkably, 
Fig.~\ref{fig:Delta}(c)
shows a second set of 
boundary scaling exponents $\Delta^{(1,\mathbb{Z}^{\ }_{2})}_q$
in the network model, which is markedly different
from the first set of boundary scaling 
exponents $\Delta^{(1,\mathrm{O})}_q$ in Fig.~\ref{fig:Delta}(b).
To sum up, the three distinct sets of scaling exponents
$\Delta^{(2)}_{q}$,
$\Delta^{(1,\mathrm{O})}_{q}$,
and
$\Delta^{(1,\mathbb{Z}^{\ }_{2})}_{q}$
are identified from 
Figs.~\ref{fig:Delta}(a), \ref{fig:Delta}(b),
and \ref{fig:Delta}(c), respectively.

The emerging picture is that
the network and the SU(2) models
at criticality share common
bulk scaling exponents $\Delta^{(2)}_{q}$
and a set of boundary scaling exponents $\Delta^{(1,\mathrm{O})}_q$
at an ordinary boundary that does not support a helical edge mode
in the neighboring insulating phase.
Only the network model has one more set of boundary scaling
exponents $\Delta^{(1,\mathbb{Z}^{\ }_{2})}_q$ at a boundary
that has a helical edge mode in the insulating phase.

\subsection{
$f(\alpha)$ spectra
           }

There is yet another set of scaling exponents 
$f^{(\kappa)}(\alpha^{(\kappa)})$
that we have calculated for the SU(2) and network models.
These exponents are calculated numerically from the scaling ansatz
\begin{subequations}
\label{eq: def alpha f alpha}
\begin{equation}
\frac{
\overline{
\overline{
\left|\Psi(\bm{r})\right|^{2q}
\ln
\left|\Psi(\bm{r})\right|^{2}
         }
         }
     }
     {
\overline{
\overline{
\left|\Psi(\bm{r})\right|^{2q}
         }
         }
     }
\sim
-
\alpha^{(\kappa)}_{q}
\ln L
\label{eq:def alpha}
\end{equation}
and
\begin{equation}
\ln
\overline{
\overline{
\left|\Psi(\bm{r})\right|^{2q}
         }
         }
\sim
\left[
f^{(\kappa)}(\alpha^{(\kappa)}_{q})
-
\alpha^{(\kappa)}_{q}q
-
d^{(\kappa)}
\right]
\ln L,
\label{eq:def f of alpha}
\end{equation}
\end{subequations}
where
$d^{(\kappa)}=2$ and $1$
for the bulk and boundary exponents, respectively.
It can be shown that the Legendre transform%
~\cite{Subramaniam06,Mildenberger07b,Obuse07a,Evers08,Obuse08}
\begin{subequations}
\begin{equation}
f^{(\kappa)}(\alpha^{(\kappa)})\equiv
(\alpha^{(\kappa)}-2)q
-
\Delta^{(\kappa)}_{q}
+
d^{(\kappa)},
\label{eq:leg f of alpha}
\end{equation}
where $q$ is a function of $\alpha^{(\kappa)}$ 
obtained from inverting
\begin{equation}
\alpha^{(\kappa)}(q)-2\equiv
d\Delta^{(\kappa)}_{q}/dq,
\label{eq:leg alpha}
\end{equation}
\end{subequations}
relates the scaling exponents
(\ref{eq: def Deltaq}) 
and
(\ref{eq: def alpha f alpha}).
The number $f^{(\kappa)}(\alpha^{(\kappa)})$ is
the fractal (i.e., Hausdorff\cite{Falconer90}) 
dimension of the set of points
$\bm{r}$ such that $|\Psi(\bm{r})|^2$ 
scales as $L^{-\alpha^{(\kappa)}}$.

\begin{figure}[t]
\centering
\includegraphics[width=0.45\textwidth]{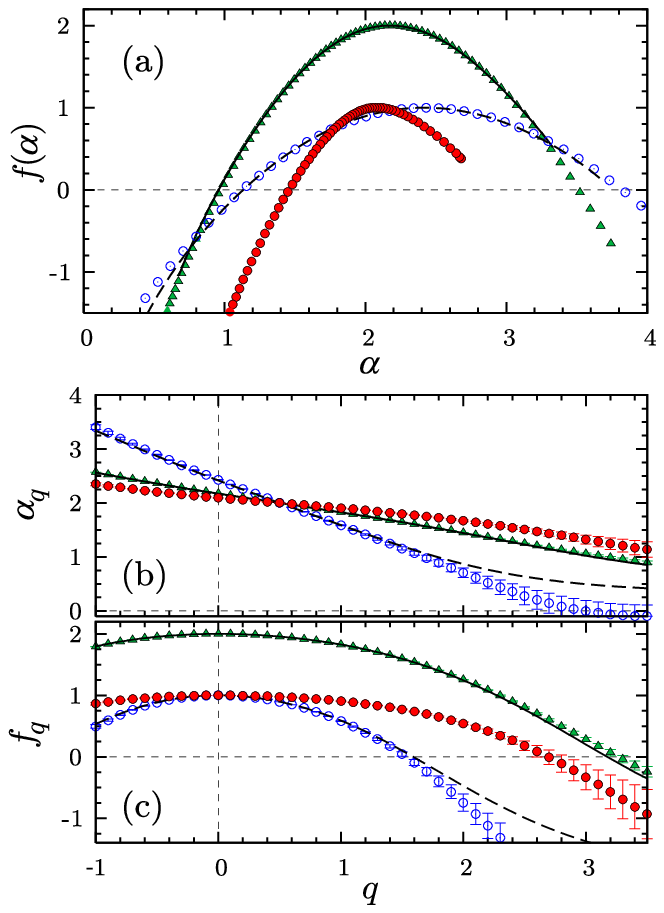}
\caption{
(Color online)
(a) Dependence on $\alpha^{(\kappa)}$ of the
multifractal spectra $f^{(\kappa)}$
in the bulk $(\kappa)=(2)$ (\textcolor[rgb]{0,0.7,0.25}{$\blacktriangle$}),
at the $\textsf{S}$ boundaries
$(\kappa)=(1,\mathrm{O})$
($\textcolor{blue}{\odot}$),
and at the $\textsf{S}'$ boundaries
$(\kappa)=(1,\mathbb{Z}^{\ }_{2})$
($\textcolor{red}{\bullet}$),
at $X=X^{\ }_{l}$.
Dependencies on $q$ of
$\alpha^{\ }_{q}$ and $f^{\ }_{q}$ are shown in (b) and (c),
respectively, with the same symbols as in (a).
The bulk and boundary multifractal spectra
for the SU(2) model defined in Ref.%
~\onlinecite{Asada02}
are shown by solid and dashed curves, respectively.
        } 
\label{fig:falpha}
\end{figure}

The dependence on $\alpha^{(\kappa)}$ of
$f^{(\kappa)}$ 
for the critical network model at $X=X^{\ }_{l}$
in the torus,
$(\kappa)=(2)$,
in the cylinder geometry with $\textsf{S}$ boundaries
[Fig.~\ref{fig:3Q1D geo}(b)], 
$(\kappa)=(1,\mathrm{O})$,
and in the cylinder geometry with $\textsf{S}'$ boundaries
[Fig.~\ref{fig:3Q1D geo}(a)],
$(\kappa)=(1,\mathbb{Z}^{\ }_{2})$,
are shown in
Fig.~\ref{fig:falpha}(a),
which are obtained by combining
$\alpha_q^{(\kappa)}$ in Fig.~\ref{fig:falpha}(b)
and $f^{(\kappa)}_q$ in Fig.~\ref{fig:falpha}(c).
We see that there are three distinct multifractal spectra
at the critical point $X^{\ }_{l}$:
$f^{(2)}$ for the bulk,
$f^{(1,\mathrm{O})}$ for the $\textsf{S}$ cylindrical geometry,
and $f^{(1,\mathbb{Z}^{\ }_{2})}$
for the $\textsf{S}'$ cylindrical geometry.
They are compared with the multifractal spectra for the SU(2) model
in the torus and cylinder geometry obtained in Ref.~\onlinecite{Obuse07a}.
Within their error bars, they agree with
$f^{(2)}$ and $f^{(1,\mathrm{O})}$, respectively.
The exponent $\alpha^{(\kappa)}_q$ at
$q=0$ is closely related
to Eq.~(\ref{eq: def <ln|Psi|2>}) 
with $x$ chosen in the bulk or on a boundary. 
For the network model we find the values
\begin{subequations}
\begin{eqnarray}
&&
\alpha^{(2)}_{0}=2.174\pm0.001,
\label{eq: alpha(2)0}
\\
&&
\alpha^{(1,\mathrm{O})}_{0}=2.420\pm0.005,
\label{eq: alpha(1,S)0}
\\
&&
\alpha^{(1,\mathbb{Z}^{\ }_{2})}_{0}=2.086\pm0.015.
\label{eq: alpha(1,S')0}
\end{eqnarray}
\end{subequations}
The inequality
$\alpha^{(1,\mathbb{Z}^{\ }_{2})}_{0}<
\alpha^{(2)}_0<
\alpha^{(1,\mathrm{O})}_{0}$
is consistent with the $x$ dependence seen at the boundaries 
in Fig.~\ref{fig:WFA},
as wave functions near $\mathsf{S}'$ boundary in 
the $\mathsf{S}'$ cylindrical geometry are expected to be more
extended because of the existence of
edge modes in the insulating side.
We note that the error bars are an order of
magnitude larger for 
$\alpha^{(1,\mathbb{Z}^{\ }_{2})}_{0}$
because of the presence of larger finite-size corrections.
Within their error bars,
the values~(\ref{eq: alpha(2)0})
and (\ref{eq: alpha(1,S)0})
agree with the ones
for the SU(2) model in the bulk and boundaries,%
\cite{Obuse07a,Mildenberger07a}
respectively.

It is worth mentioning that the multifractal analysis performed
here involves extracting scaling exponents 
after performing the disorder averaging.
Extracting scaling exponents before performing the disorder
averaging yields typical scaling exponents.
Typical scaling exponents need not be identical with average
scaling exponents calculated here.%
~\cite{Mudry96,Chamon96,Mudry03}
The average scaling exponents $\Delta^{(\kappa)}_{q}$
are expected to differ from the typical ones
for any values of $q$ such that $f^{(\alpha)}(\alpha^{(\kappa)})$
is negative;\cite{Mirlin00} see Fig.~\ref{fig:falpha}(c).
For this range of $q$ rare events dominate
the calculation of $\Delta^{(\kappa)}_{q}$
as is evidenced by the larger error bars in Fig.~\ref{fig:Delta}.
This explains the systematic deviations from the mirror symmetry%
~\cite{Mirlin06} 
about $q=1/2$ of
$\Delta^{(\kappa)}_{q}/[q(q-1)]$
for large $q$ in Fig.~\ref{fig:Delta}.

The rational for studying these average scaling exponents is
that they are expected to be the 
scaling dimensions of some primary
operators representing moments of wave function amplitudes
in an underlying two-dimensional conformal field theory.\cite{Duplantier91}
Knowing them constrains the possible field theories
that can encode critical properties of an Anderson transition.

\section{
Corner multifractality
        }
\label{sec:Corner multifractality}

So far we have always considered geometries of the network model
with boundaries of the same type, as in 
Figs.~\ref{fig:3Q1D geo}(a) 
or~\ref{fig:3Q1D geo}(b).
We are now going to investigate the case of mixed boundaries.

We begin with the network model in the
quasi-one-dimensional geometry (quantum wire geometry) from
Fig.~\ref{fig:3Q1D geo}(c)
for which we show that the Landauer conductance obeys a statistical
distribution that differs from that of an ordinary symplectic quantum wire, 
i.e., a symplectic quantum wire in a geometry compatible 
with transverse PBC.
A physical realization of this case is given by metallic carbon
nanotubes with spatially smooth disorder potential.
As shown by Ando and Suzuura,\cite{Ando02} in the absence of intervalley
scattering by disorder, conduction of electronic states near a Fermi point
is described by transfer matrices in the symplectic class with
an odd number of conduction channels.
A mathematical treatment of such conduction process was initiated
by Zirnbauer\cite{Zirnbauer92} and
further developed by Takane.\cite{Takane04}

\begin{figure}[t]
\centering
\includegraphics[width=0.45\textwidth]{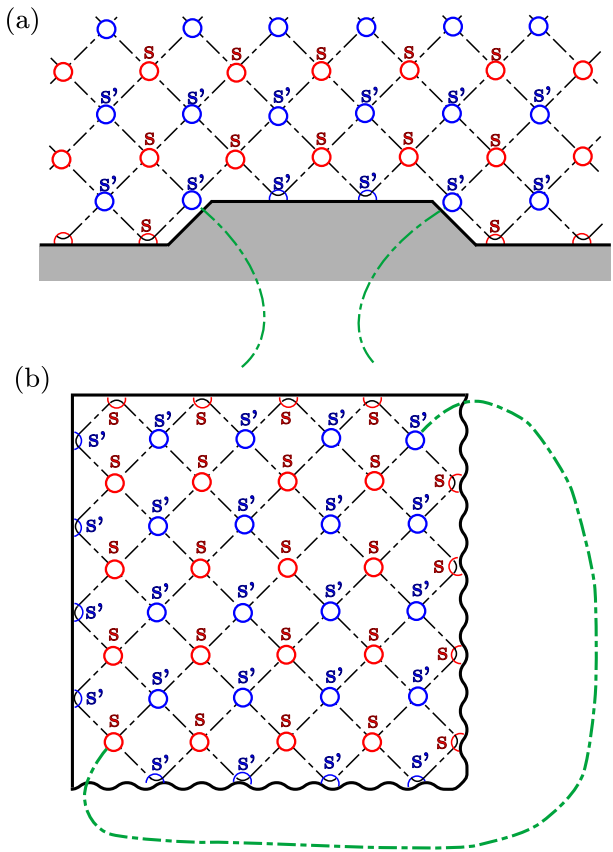}
\caption{
(Color online)
(a)
A semi-infinite geometry with
two point contacts (green dot-dashed curves)
attached at the two interfaces
between different types of boundaries.
Each line or curve represents a Kramers doublet.
(b) 
Closed network with mixed boundaries.
The thick wavy and solid lines on the edges represent
two different types of boundaries, with and without
a helical edge mode at $X>X^{\ }_{l}$, respectively.
        } 
\label{fig:mixed boundaries}
\end{figure}

We then turn our attention to a semi-infinite geometry with
two point contacts to ideal reservoirs attached at the interface
between the boundary of type \textsf{S} and of type \textsf{S}'
as shown in Fig.~\ref{fig:mixed boundaries}(a).
After tuning the network model to criticality,
we shall see that this amounts to studying corner multifractal scaling
exponents whose values differ from those taken by the corner multifractal
scaling at the corner along a homogeneous segment of a boundary,
see Fig.~\ref{fig:mixed boundaries}(b).

Common to both examples of mixed boundaries
is the existence of a single perfectly conducting channel.

\subsection{
Quasi-one-dimensional wire with mixed transverse open boundary conditions
           }

In the quasi-one-dimensional wire
depicted in Fig.~\ref{fig:3Q1D geo}(c), 
whereby the length of the disordered region
$N$ is larger than the mean free path $\ell\gg1$,
one transmission eigenvalue is always unity
for any disorder configuration
while the other $2M$ eigenvalues are exponentially small with $N$.
This follows from applying
the results from Refs.~\onlinecite{Ando02} and \onlinecite{Takane04}.
The same is true of the thick quantum wire limit
$N,M\to\infty$ with the ratio $M\ell/N$ held fixed,
as it is now the analysis from Ref.~\onlinecite{Zirnbauer92}
that can be borrowed.

The persistence of the unit eigenvalue of the transmission matrix
is unique to the network model with mixed boundary conditions.
This remarkable property can be understood physically as follows.
Because of the mixed boundaries, one and only one edge
supports a single Kramers doublet of edge states.
This Kramers doublet of edge states is free to propagate
along the edge and thus provide a single perfectly conducting channel.
This physics is indeed realized with helical edge modes of
a two-dimensional quantum spin Hall insulator
and a metallic carbon nanotube in the absence of intervalley
scattering. Mathematically,
the transfer matrix $\mathcal{M}$ for the geometry
of Fig.~\ref{fig:3Q1D geo}(c)
is a member of the group $\mathrm{SO}^*(4M+2)$.\cite{Obuse07b}
The eigenvalues of $\ln(\mathcal{M}\mathcal{M}^\dagger)$ are symmetrically
distributed around 0 and denoted by $\pm 2x^{\ }_{i}$
($0\le x^{\ }_{1}\le x^{\ }_{2}\le\cdots\le x^{\ }_{2M+1}$).
In turn, the transmission eigenvalues $T^{\ }_{i}$ are written as
$T^{\ }_{i}=1/\cosh^2x^{\ }_{i}$ ($i=1,2,\ldots,2M+1$).
Because of the Kramers degeneracy, 
the eigenvalues of $\ln(\mathcal{M}\mathcal{M}^\dagger)$
are two-fold degenerate.
This leads to the conclusion that there must be a null eigenvalue
$x^{\ }_{1}=0$, thereby $T^{\ }_{1}=1$.

\subsection{
Two point contacts on a boundary of mixed type
           }

Next, we consider the case of a semi-infinite network model
with a boundary that is mixed, whereby
it is necessary to attach two point contacts (leads)
at the two interfaces between
the boundary nodes of type \textsf{S} and \textsf{S}'
in order to maintain TRS. This setup is shown in 
Fig.~\ref{fig:mixed boundaries}(a).
The scattering matrix that relates incoming to outgoing waves from the
leads is then a $2\times2$ matrix.
The constraint that it belongs to the symplectic symmetry class
forces this matrix to be proportional to the unit $2\times2$ matrix up
to an overall (random) phase. Hence, the two-point conductance is always
unity however far separated the two point contacts are.
An incoming Kramers doublet is transmitted with probability one
through the disordered region in 
Fig.~\ref{fig:mixed boundaries}(b).
This result may be related to the presence
of a perfect transmission channel in the geometry of
Fig.~\ref{fig:3Q1D geo}(c)
by thinking of a conformal mapping
that transforms a half plane with mixed boundaries to
an infinite strip with mixed boundaries.

Before proceeding with the case at hand, we recall that 
it is expected on general grounds that the moments of the 
two-point conductance
in a network model at criticality,
when the point contacts are far apart,
decay as power laws with scaling exponents proportional to
the scaling exponents $\Delta^{(\kappa)}_{q}$.%
~\cite{Klesse01}
Consequently, after tuning the semi-infinite network model to criticality,
the zero-dimensional boundary scaling exponents
for the moments of $|\Psi(\bm{r})|^2$ with $\bm{r}$ at
an interface of different types of boundaries
\begin{equation}
\Delta^{(0,\mathbb{Z}^{\ }_{2}|\mathrm{O})}_{q}=0
\label{eq: Delta(0,T|O)=0}
\end{equation}
emerge as a signature of the nontrivial topological nature of 
the insulating side at the Anderson transition.
We conjecture that, if a description of 
the critical point exists in terms of a conformal field theory, 
the exponents~(\ref{eq: Delta(0,T|O)=0}) might then be obtained 
from the correlation functions between the moments of the local operator
encoding the local density of states
and additional insertions of a boundary condition changing
operator.\cite{Cardy1989}

The zero-dimensional multifractal scaling
exponents~(\ref{eq: Delta(0,T|O)=0})
are different from the corner
multifractal scaling exponents
which characterize scaling of the moments
of wave functions at a corner with the boundary of a given type,
such as the upper left corner and the lower right corner in 
Fig.~\ref{fig:mixed boundaries}(b).
These corner multifractal exponents read
\begin{equation}
\Delta^{(0,\mathrm{O})}_{q}=
2
\Delta^{(1,\mathrm{O})}_{q}
\label{eq: Delta(0,O)}
\end{equation}
or
\begin{equation}
\Delta^{(0,\mathbb{Z}^{\ }_{2})}_{q}=
2
\Delta^{(1,\mathbb{Z}^{\ }_{2})}_{q},
\label{eq: Delta(0,T)}
\end{equation}
depending on the type of the boundary and the critical point
to which the network model has been tuned.
Here, Eqs.~(\ref{eq: Delta(0,O)}) and (\ref{eq: Delta(0,T)})
follow from a general relation based on 
two-dimensional conformal mappings
that relates the scaling exponents~(\ref{eq: def Deltaq})
with $\bm{r}$ 
at the corner between two straight boundaries
meeting with the angle $\pi/2$
and the scaling exponents~(\ref{eq: def Deltaq})
with $\bm{r}$ along a straight boundary.\cite{Obuse07a}

\section{
Conclusions
        }
\label{sec: Conclusions} 

We have shown that the boundary multifractal spectra for the critical
normalized wave functions are sensitive to the choice
of boundary conditions
at the Anderson transition in the two-dimensional symplectic class.
This is the first example where boundary multifractal exponents are
calculated under different boundary conditions in the problem of
Anderson localization-delocalization transition.
It would be interesting to look for other examples of disorder-induced
continuous phase transitions at which boundary critical properties
depend on boundary conditions but bulk critical properties do not.

We conjecture that the two-dimensional conformal theory describing
the Anderson transition in the two-dimensional symplectic
universality class, if it exists, should be compatible with two distinct
conformally invariant boundary conditions on the boundary of a half-plane,
thereby yielding two distinct sets of boundary multifractal exponents
$\Delta^{(1,\mathrm{O})}_q$ and
$\Delta^{(1,\mathbb{Z}^{\ }_{2})}_q$.
The recent observation of the QSH effect in HgTe/(Hg,Cd)Te quantum wells%
~\cite{Konig07}
suggests that it might be possible to
probe experimentally the Kramers degenerate edge states 
at criticality and the corresponding boundary multifractal spectra.

\section*{Acknowledgments}

This work was supported by the Next Generation Super Computing Project,
Nanoscience Program, MEXT, Japan and by the National Science Foundation
under Grant No.\ PHY05-51164. 
Numerical calculations have been mainly performed on the RIKEN Super
Combined Cluster System.


\begin{thebibliography}{40}
\bibitem{note on the opposite limit}
A similar analysis can be done in the opposite limit of $X\to0$,
for which the topological nature of
the insulating state can be deduced by considering the limit of
small $\theta(>0)$.  This effectively amounts to 
the interchange of the nodes \textsf{S} and $\textsf{S}'$.
(Taking the small-$\theta$ limit is justified by the phase diagram
in Fig.~11 of Ref.~\onlinecite{Obuse07b}.)
By doing so, the conclusion as to whether a helical edge state
is present on upper/lower boundaries is reversed 
compared to the conclusion for the $X\to\infty$ limit.
\bibitem{footnote:Kramers degeneracy Lyapunov} 
Lyapunov exponents are two-fold degenerate 
due to the Kramers degeneracy
in the geometries of Figs.~\ref{fig:3Q1D geo}(a) and
\ref{fig:3Q1D geo}(b),
for which the transfer matrices are $4M\times4M$ matrices.
\end{thebibliography}
\end{document}